\documentclass[aps,prd,tightenlines,superscriptaddress,nofootinbib,twocolumn]{revtex4}

\usepackage{graphicx}
\usepackage{amsmath}
\usepackage{color}
\usepackage{natbib}
\usepackage{epstopdf}

\def\be{\begin{equation}}
\def\ee{\end{equation}}
\def\ben{$$}
\def\een{$$}
\def\bea{\begin{eqnarray}}
\def\eea{\end{eqnarray}}
\def\bean{\begin{eqnarray*}}
\def\eean{\end{eqnarray*}}

\def\bi{\begin{itemize}}
\def\ei{\end{itemize}}
\def\ben{\begin{enumerate}}
\def\een{\end{enumerate}}

\begin{document}

\title{Identifying the Host Galaxy of Gravitational Wave Signals} 

\author{Laura K.~Nuttall, Patrick J.~Sutton}
\affiliation{School of Physics and Astronomy, Cardiff University, Cardiff, United Kingdom, CF24 3AA}

\begin{abstract}
One of the goals of the current LIGO-GEO-Virgo science run is to identify 
transient gravitational wave (GW) signals in near real time to allow 
follow-up electromagnetic (EM) observations. An EM counterpart could increase 
the confidence of the GW detection and provide insight into the nature of the 
source. Current GW-EM campaigns target potential host galaxies based on overlap 
with the GW sky error box. We propose a new statistic to identify the most likely 
host galaxy, ranking galaxies based on their position, distance, and luminosity. 
We test our statistic with Monte Carlo simulations of GWs produced by 
coalescing binaries of neutron stars (NS) and black holes (BH), one 
of the most promising sources for ground-based GW detectors. 
Considering signals accessible to current detectors, we find that when 
imaging a single galaxy, our statistic correctly identifies the true host 
$\sim$20\% to $\sim$50\% of the time, depending on the masses of the binary 
components.  With five narrow-field images the probability of imaging 
the true host increases to $\sim$50\% to $\sim$80\%.
When collectively imaging groups of galaxies using large field-of-view 
telescopes, the probability improves to $\sim$30\% to $\sim$60\% for 
a single image and to $\sim$70\% to $\sim$90\% for five images.  
For the advanced generation of detectors (c.~2015$+$), and considering 
binaries within 100 Mpc (the reach of the galaxy catalogue used), the 
probability is $\sim$40\% for one narrow-field image,
$\sim$75\% for five narrow-field images, $\sim$65\% for one wide-field image, 
and $\sim$95\% for five wide-field images, irrespective of binary type.
\end{abstract}



\maketitle

\section{Introduction} 
\label{sec:intro} 
The next decade should see the first direct detection of gravitational waves 
with the global network of GW detectors. The two LIGO 
detectors \cite{Abbott:2007kv,Smith:2009bx} 
are located in Louisiana and 
Washington state, USA, the joint French-Italian Virgo detector \cite{virgo08} 
in Pisa, Italy and the German-British detector GEO-600 \cite{geo08} in 
Hannover, Germany. The 2005-2007 science run of LIGO-GEO-Virgo saw the two 
LIGO detectors take data at design sensitivity.
Since 2009 LIGO and Virgo have been taking data with improved sensitivities, 
and GEO also recommenced data taking in 2010.
It is expected that in the advanced detector era (c.~2015$+$) LIGO and Virgo 
will operate at sensitivities more than 10 times greater than initial LIGO, 
thereby increasing the volume of monitored universe by more than a factor of 
1000 \cite{aligo,avirgo}. In addition new detectors, such as LCGT 
in Japan \cite{kuroda:2010} and AIGO 
in Australia \cite{Blair:2008zzc}, are being planned.

Electromagnetic identification of a GW might not only confirm a GW detection, but also improve 
parameter extraction, and, by independently identifying the source's position 
and time, lower the signal-to-noise ratio (SNR) required for a confident 
detection.  Joint GW-EM observations could also address specific questions such 
as the nature of short hard $\gamma$-ray bursts and allowing a more precise 
measurement of $H_{0}$; see Bloom et al.~\cite{Bloom:2009vx} for an overview.

GW detectors are non-imaging detectors with a large field of view; their  
antenna response is greater than half-maximum over 65\% of the sky.
Source localization for short-lived signals therefore requires multiple detectors, in order to use the 
measured time delay between detectors as well as the amplitude of the measured 
signal in each detector to triangulate a sky location.  Several methods of 
localization have been investigated 
\cite{Guersel:1989th,Wen:2005ui,Cavalier:2006rz,Rakhmanov:2006qm,Acernese:2007zza,Searle:2007uv,Wen:2008zzb,Markowitz:2008zj,Fairhurst:2009tc,Searle:2008ap,Wen:2010cr}. 
Fairhurst \cite{Fairhurst:2009tc} gives the following approximation for the timing 
accuracy of a GW signal:
\begin{equation}
\label{timing}
 \sigma_{t} \sim \frac{1}{2\pi \sigma_{f} \rho} \, ,
\end{equation}
where $\sigma_{f}$ is the effective bandwidth of the signal and $\rho$ is the 
SNR. For nominal values $\sigma_f = 100$ Hz and $\rho=8$, timing accuracies are 
on the order of 0.1 ms.  This can be compared to the light travel time 
between detectors, 10 -- 30 ms for the LIGO-Virgo network.  
For example, for a binary coalescence signal at the threshold of detectability,
Fairhurst \cite{Fairhurst:2009tc} estimates a best-case localization of 20 deg$^{2}$ 
(90\% containment), and a typical localization of twice this.

The LOOC UP (Locating and Observing Optical Counterparts to Unmodeled Pulses) 
project \cite{Kanner:2008zh} consists of reconstructing the sky position of 
candidate GW signals and making prompt follow-up observations using wide 
field-of-view cameras. However, given the large sky error box associated with GW signals, 
identifying the source of a GW signal is not trivial; over 100 
galaxies can be found within an error box out to 100 Mpc. We therefore 
present a ranking statistic for identifying the most likely host galaxy based on 
galaxy distance and luminosity and the sky position error box. 
Using Monte Carlo simulations of GW signals, we demonstrate that this ranking 
statistic can correctly identify the host galaxy for a significant fraction of 
GW signals detectable by the initial and advanced LIGO-Virgo networks.

\section{Sources of GWs and Optical transients}
\label{sec:sources}

Due to the strongly relativistic nature of GW sources, many systems that would 
produce detectable GW emission should also be observable electromagnetically.
Cutler \& Thorne \cite{Cutler:2002me} provide a brief review of sources of GWs observable by 
ground-based detectors.  Mergers of binary neutron stars (NS-NS) or binaries consisting of a neutron star and a stellar mass black hole (NS-BH) are the best understood in terms of GW range 
and expected rate \cite{Abadie:2010cf}, and are the most likely sources for 
producing both detectable GW signals and optical transients.  They are also 
the favoured progenitor model for short $\gamma$-ray bursts \cite{nakar:2007} 
\footnote{The $\gamma$ emission might itself be used to identify the host galaxy for 
those cases where the emission is beamed towards us.}. 
These events will emit a 
significant proportion of their binding energy in GWs at frequencies to which 
the current and next generation of GW detectors are sensitive. 
The distance to which a GW signal can be detected depends on the masses of 
the binary components. Assuming fiducial masses of 
1.4 $M_{\odot}$ for neutron stars and 10 $M_{\odot}$ for black holes, 
the current LIGO observatories can detect NS-NS binary systems with $\rho \ge 8$ 
out to a maximum distance of approximately 30 Mpc, and NS-BH systems out to 65 Mpc. 
With this sensitivity, Abadie et al.~\cite{Abadie:2010cf} estimate the most likely rate of 
detectable signals at $\sim$0.02 yr$^{-1}$ for NS-NS and $\sim$0.004 yr$^{-1}$ 
for NS-BH systems.
Estimates of the optical emission due to the radioactive decay of heavy 
elements synthesized in the merger ejecta predict a peak magnitude of 
approximately 20 or brighter at 65 Mpc 
\cite{Li:1998bw,rosswog:2005,Metzger:2010sy}. 
For advanced LIGO (c.~2015$+$) the GW range increases to approximately 450 Mpc 
for NS-NS systems and 930 Mpc for NS-BH systems, with most likely rate estimates 
of $\sim$40 yr$^{-1}$ and $\sim$10 yr$^{-1}$ respectively.  
This could require optical imaging to magnitude 25-26 for the most distant sources. 
Restricting to sources within 100 Mpc (the reach of our galaxy catalogue, discussed below), 
the expected detection rates are $\sim$5 yr$^{-1}$ for NS-NS and 
$\sim$0.2 yr$^{-1}$ for NS-BH \footnote{The fraction of detections 
that occur within 100 Mpc may be estimated as $(100~\textrm{Mpc}/R_\textrm{SM})^3$, 
where $R_\textrm{SM}$ is the ``sensemon range'' \cite{Abadie:2010cf,Finn:1992xs}. 
$R_\textrm{SM}$ is the radius of a sphere whose volume is the effective 
volume in which a source can be detected, taking into account all possible 
sky locations and binary orientations. $R_\textrm{SM}$ is a factor of 2.26 
smaller than the maximum detection distance.}.

Another possible EM-GW source is core-collapse supernovae; however, 
even for the advanced detectors, GW emission from these systems is likely 
to be detectable only for supernovae occurring within our galaxy 
\cite{Ott:2008wt}, for which the rate is $\sim$0.02 yr$^{-1}$.  
We therefore focus our attention on NS-NS and NS-BH systems.

\section{GW Catalogue}
\label{sec:catalogue}

We simulate GW signals coming from known external 
galaxies, using the Gravitational Wave Galaxy Catalogue of 
White et al.~\cite{White:2011qf}.  This catalogue contains approximately 53,000 galaxies out 
to a distance of 100 Mpc.  There are $22,000$ galaxies within 65 Mpc, 
the maximum distance to which a 1.4-10.0 $M_{\odot}$ NS-BH system can be 
detected with $\rho\ge8$ by initial LIGO, and $7300$ galaxies 
within 30 Mpc, the maximum distance for a NS-NS binary.  
White et al.~estimate the catalogue to 
have a completeness of 60\% to 100 Mpc, 75\% to 50 Mpc, and a 
completeness consistent with 100\% out to 40 Mpc.  
Approximately 50\% of the galaxies have a defined type in the de Vaucouleurs classification \cite{De:1959cm}; these account for $80\%$ of the total luminosity in the catalogue.

\section{Ranking Statistic}
\label{sec:statistic}

A LIGO-Virgo GW error box can contain over one hundred galaxies 
out to 100 Mpc.  Imaging all to search for an EM counterpart will likely 
be impractical.  This motivates considering ways to 
rank the galaxies by their likelihood of hosting the source of the observed GW event.

We expect a nearby galaxy to be more likely \emph{a priori} to be the host of a detectable GW signal source than a more distant galaxy.  Furthermore, larger galaxies 
contain more potential sources. We therefore propose to rank each galaxy 
as the possible host for a GW signal by the following statistic:
\begin{equation}
\label{rank}
 R = e^{-\frac{\chi^{2}}{2}}\frac{L}{d^{\alpha}} \, .
\end{equation}
Here $L$ is the luminosity of the putative host galaxy, $d$ is the distance to 
the galaxy, $\alpha$ is a constant, and $\chi^{2}$ is the chi-squared match between the measured and 
predicted time of arrival of the signal in each detector \cite{Cavalier:2006rz}, given by
\begin{equation}
\label{chi}
 \chi^{2} = \sum_{i}\frac{(t_{i}-p_{i})^{2}}{\sigma^{2}_{i}} \, .
\end{equation}
Here $\sigma_{i}$ is the timing uncertainty in each detector, $t_{i}$ is the 
measured arrival time, $p_{i}$ is the predicted arrival time based on the sky 
direction of the putative host galaxy, and the sum is taken over all detectors. 
We include $\exp{(-\chi^{2}/2)}$ in our ranking statistic as this is the 
likelihood associated with a Gaussian timing error in each detector.
It determines which galaxies have sky positions consistent with the observed 
time delays between detectors; i.e., it represents the GW triangulation error box\footnote{
For the LIGO-Virgo network that we will simulate, the $\chi^2$ sky 
map is mirror-symmetric through the plane of the detectors, thus usually yielding 
two error boxes.  In principle, the measured signal SNRs can be used to break this 
degeneracy and determine which box contains the correct sky location.  For our 
tests, we use both boxes.  Therefore, a more sophisticated GW analysis than that 
assumed here may reduce the number of galaxies that need to be imaged by up to a 
factor of 2.}.

We scale $R$ with luminosity because we assume the luminosity of each galaxy 
to be approximately proportional to the number of sources within it. 
The $d^{-\alpha}$ factor favours intrinsically weak signals from nearby 
galaxies as being more likely than strong signals from distant galaxies. 
More generally, if we assume the rate of GW events of intrinsic amplitude 
$h_0$ within each galaxy to be of the form
\begin{equation}
\frac{dN}{dh_0}\sim h_0^{-\alpha} \, ,  
\end{equation}
then, since the received amplitude $h$ is $h \propto h_0 d^{-1}$, the correct distance weighting is $d^{-\alpha+1}$.  
In our simulations we test $\alpha = 1,2,3$. 
We find $\alpha=2$ gives marginally better performance for the initial LIGO detectors, 
and $\alpha=1$ the best for advanced LIGO.   However, the variation in the probability of 
identifying the host galaxy is only a few percent; we conclude that our ranking is 
not sensitive to the precise distance weighting used.

For comparison, we also test ranking based purely on the error box, 
with no luminosity or distance weighting:
\begin{equation}
\label{noweight}
R=e^{-\frac{\chi^{2}}{2}}.
\end{equation}
This statistic is poor at identifying the host galaxy; the probability 
of correct identification is a factor of 2-4 lower (depending on binary 
mass) than when including the $L/d$ weighting.

\section{Simulations}
\label{sec:simulations}

To evaluate how well our ranking statistic identifies the true host galaxy of a 
GW signal, we simulate how GWs will appear in a realistic search. We consider 
inspiralling NS-NS and NS-BH binaries. The strength of their 
GWs has a well-defined dependence on the system's mass, distance, and 
inclination of the binary orbital axis to the line of sight.  We study 3 
different mass pairs: 1.4-1.4 $M_{\odot}$ NS-NS, 1.4-5.0 $M_{\odot}$ NS-BH, 
and 1.4-10.0 $M_{\odot}$ NS-BH systems. The orientations are random and 
isotropic.  The true host galaxy is selected randomly with weight proportional to the 
galaxy luminosity and with an additional weighting based on galaxy type as 
discussed below.  

We simulate the LIGO-Hanford, LIGO-Livingston and Virgo network, assuming 
all three detectors to have sensitivity given by the initial LIGO design 
\cite{Abbott:2007kv}, or the advanced LIGO design \cite{aligo}.  
For each GW, we compute the received SNR in 
each detector based on the binary mass and distance, and the detector 
sensitivity to that sky direction and binary orientation.  We also compute the 
timing uncertainty using equation (\ref{timing}).  The measured amplitudes and times  
are ``jittered'' by additive Gaussian errors to simulate the detector noise 
background.  To be considered detected, a GW 
needs to have an SNR of 
$\rho\ge8$ in at least two detectors, and a quadrature-sum SNR over all three 
detectors $\ge12$.  For each Monte Carlo run we generate enough binaries to give 
approximately 800 detected signals.

While our ranking statistic (equation (\ref{rank})) treats all galaxy types equally, 
the rate of binary coalescences is likely to be different in different 
galaxy types. O'Shaughnessy et al.~\cite{O'Shaughnessy:2009ft} estimate the rate of NS-NS 
and NS-BH mergers in elliptical and spiral galaxies for a large range of 
plausible binary evolution scenarios.  They produce a total of 488  
samples of merger rates, and find the relative rate in spirals and 
ellipticals to vary widely in their models.  
We account for this uncertainty in our simulations by performing 50 separate 
Monte Carlo runs for each waveform type; in each run, the relative rate of 
mergers in spirals and ellipticals is determined by a random draw from the models by O'Shaughnessy et al.
We treat lenticular galaxies as equivalent to ellipticals and irregular galaxies 
as spirals for these simulations.  For those galaxies without 
a specified type, one is assigned randomly in proportion to the number of 
galaxies of each type in the catalogue.  In all, $70\%$ of the galaxies are 
treated as spiral, and $30\%$ as elliptical galaxies.

Finally, to simulate the effect of measurement errors in the galaxy catalogue 
we also 
jitter the luminosity and distance of each galaxy by a random amount consistent 
with the stated uncertainties. This is done by creating a second copy of the 
galaxy catalogue and using this jittered catalogue for signal generation 
(keeping the original catalogue for ranking). 

\begin{figure}
\includegraphics[width=0.45\textwidth]{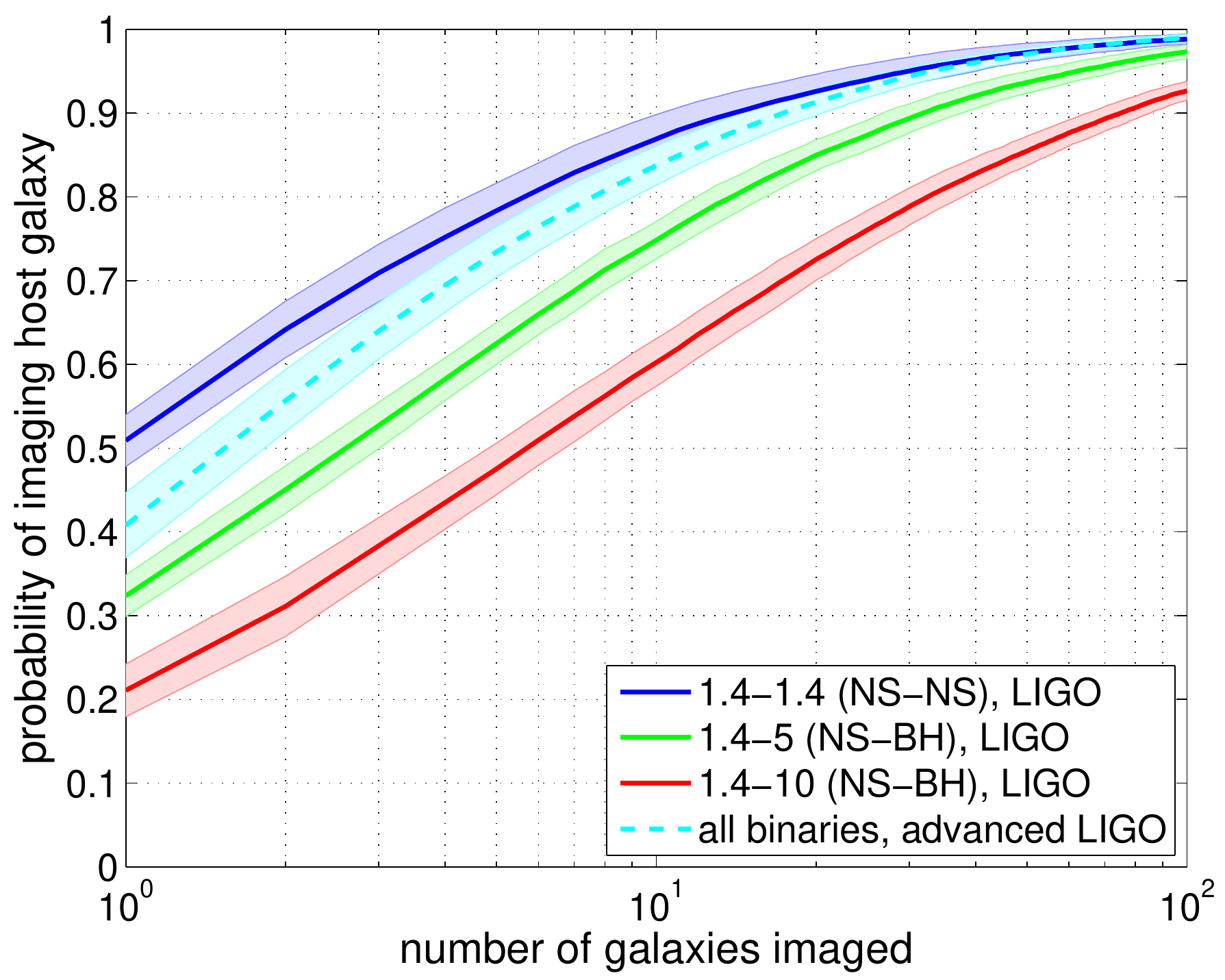}
\caption{\label{fig:proball} Narrow field of view case. The probability of 
imaging the true host galaxy for each type of binary system versus the 
number of images taken. The shaded regions denote the 1-sigma uncertainty in 
the probability estimate.}
\end{figure}

After the GW signals are generated, we compute the $\chi^{2}$ match (equation 
(\ref{chi})) between the predicted and the measured GW arrival 
time at each detector.  We then rank all the galaxies as potential hosts for 
each GW using equation (\ref{rank}).  The distribution of ranks assigned to the 
true host galaxy for each GW then tells us the probability of observing the true 
host as a function of the number of galaxies imaged.  This probability is shown 
in Figure \ref{fig:proball}.  We find that for a narrow field-of-view telescope 
($O$(10) arcmin, sufficient to image one galaxy at 10 Mpc) 
the probability of the true host being the top-ranked galaxy  
is $50\pm3\%$ for a 1.4-1.4 $M_{\odot}$ NS-NS system, 
$32\pm2\%$ for a 1.4-5.0 $M_{\odot}$ NS-BH, and 
$21\pm3\%$ for a 1.4-10.0 $M_{\odot}$ NS-BH system. 
When imaging the 5 highest-ranked galaxies, the chances of including the true 
host increase to $78\pm3\%$, $63\pm3\%$, and $48\pm3\%$ respectively.
For the advanced LIGO detectors, and considering only binaries within 100 Mpc, 
the probabilities are approximately independent of binary type: 
$39\pm3\%$ / $43\pm4\%$ / $40\pm 3\%$ for 1 image and 
$72\pm 3\%$ / $75\pm 3\%$ / $73\pm 3\%$ for 5 images.
In each case the uncertainties are dominated by the range of possible relative 
rates for mergers in spiral versus elliptical galaxies.

We note that the success rate for initial LIGO is highest for NS-NS systems, and 
decreases with increasing binary mass.  This is due to two factors.  The effective 
bandwidth $\sigma_f$ is larger for low-mass systems, giving smaller timing 
uncertainties (see equation (\ref{timing})).  Furthermore, less massive 
binaries are detectable to smaller 
distances, hence there are fewer potential hosts for these systems, so the 
probability of imaging the true host increases.  Indeed, in the NS-NS 
simulations for current detectors, we find that 10\% of all detected signals 
are due to only 10 galaxies: PGC047885, NGC0224 (Andromeda galaxy), NGC4594 
(Sombrero galaxy), ESO468-020, NGC0253, NGC5457 (Pinwheel galaxy), 
NGC6964, PGC2802329, PGC009892 and NGC4472.  It may therefore be worthwhile 
to take reference images of these ``most promising'' galaxies {\em before} 
the GW search is performed, to allow immediate identification of an EM 
transient when one of these galaxies is selected for follow-up of a GW 
signal.

For the advanced LIGO detectors, we find that the probability of imaging 
the true host galaxy is approximately the same for all binary types.  This 
is due to the restriction to signals originating within a fixed distance of 
100 Mpc. Higher-mass systems give larger SNR $\rho$ at a fixed distance; this 
offsets the effect of their lower effective bandwidth $\sigma_f$ in the 
timing uncertainty (equation (\ref{timing})).

\begin{figure}
\includegraphics[width=0.45\textwidth]{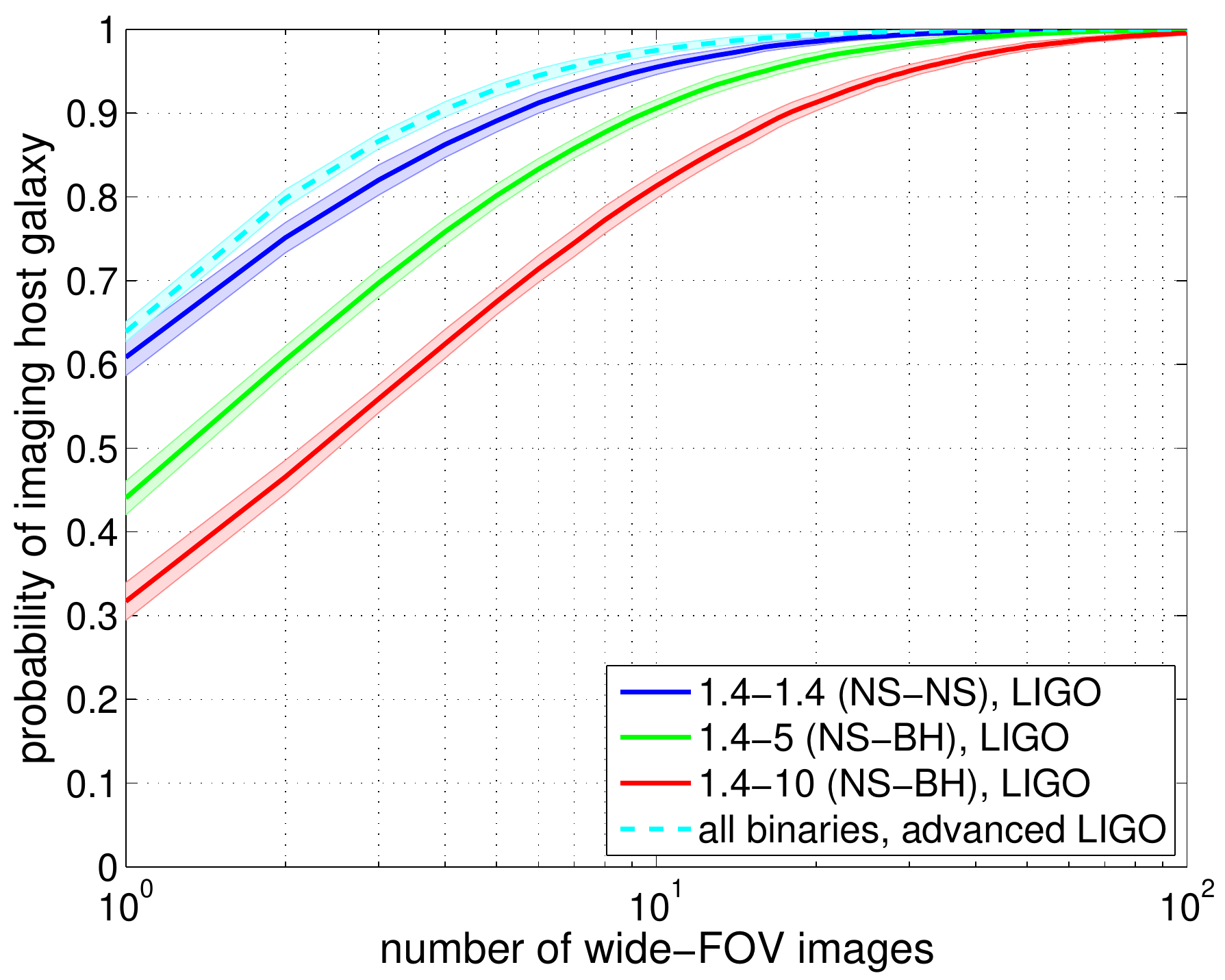}
\caption{\label{fig:proballwide} Wide field of view case. The probability of 
imaging the true host galaxy for each type of binary system versus the 
number of images taken. The shaded regions denote the 1-sigma uncertainty in 
the probability estimate.}
\end{figure}

The LOOC UP program \cite{Kanner:2008zh} is currently using wide field-of-view 
telescopes to image potential host galaxies, including TAROT \cite{Klotz:2009xd}, 
QUEST \cite{Baltay:2007kp}, and SkyMapper \cite{Keller:2007sm}, as well as 
narrow-field telescopes such as Zadko \cite{Coward:2010wi}.  
Depending on the length of exposure (between 60 s 
and 180 s) and the filter used, these telescopes have limiting magnitudes 
ranging from 17 to 22, sufficient to detect the EM emission from binary mergers 
predicted by Metzger et al.~\cite{Metzger:2010sy} to 15 -- 150 Mpc. 
The wide-field telescopes can image several square degrees at once, allowing 
multiple galaxies to be observed simultaneously and therefore increasing 
the probability of observing the true host in a given number of exposures.  
We simulate imaging with a 3-4 deg$^2$ field of view telescope by grouping 
galaxies which lie within 1 deg of one another when computing the 
probability of imaging the host.  That is, we consider the true host as 
having been imaged if it lies within 1 deg of any of the $N$ top-ranked 
galaxies, where $N$ is the number of wide-field images taken.  The results 
are shown in Figure \ref{fig:proballwide}.  We find that for initial LIGO, 
for 1.4-1.4 $M_{\odot}$ / 1.4-5.0 $M_{\odot}$ / 1.4-10.0 $M_{\odot}$ 
systems the chances of observing the true host are 
$61\pm2\%$ / $44\pm2\%$ / $32\pm 2\%$ for 1 image and 
$89\pm 1\%$ / $80\pm 1\%$ / $67\pm 2\%$ for 5 images.
These are a factor of about 1.2 better than the narrow-field-of-view 
results.
For the advanced LIGO detectors the probabilities are 
$64\pm1\%$ / $68\pm1\%$ / $64\pm 1\%$ for 1 image and 
$93\pm 1\%$ / $94\pm 1\%$ / $92\pm 1\%$ for 5 wide-field images, 
a factor of 1.3-1.5 better than in the narrow-field-of-view case.

\section{Summary and Conclusion}
\label{sec:conclusion}

We propose a ranking statistic for identifying the host galaxy of 
gravitational wave signals.  The ranking is based on the galaxy 
distance, luminosity, and overlap with the GW sky position error box.  We have 
tested the statistic by simulating GW signals from coalescing binaries 
of neutron stars and black holes and using the 
Gravitational Wave Galaxy Catalogue of White et al.~\cite{White:2010gc}.
For the current LIGO-Virgo network we find the probability of 
the true host being within the field of view of the top-ranked 
galaxy ranges from $\sim$20\% to $\sim$60\%, depending on the 
masses of the binary components and whether the observations 
are made with narrow field-of-view or wide field-of-view telescopes.
The probability of the true host being in the top 5 ranges from 
$\sim$50\% to $\sim$90\%.  
For the advanced LIGO-Virgo network, and restricting to binaries within 
100 Mpc (the range of our catalogue), the probability of 
the true host being the top-ranked ranges from $\sim$40\% to $\sim$70\%, 
and in the top 5 from $\sim$70\% to $>$90\%. 
In general our ranking statistic favours
larger, closer galaxies as the most probable hosts, and so performs best 
for GWs from nearby galaxies.

Our simulations account for uncertainties in the relative rate of mergers 
in galaxies of different types, as well as uncertainties in the 
measured properties of the galaxies (distance, luminosity, and type).  
We find these effects change the probability of imaging the true 
host by only a few percent.  We have also verified that the ranking is 
not sensitive to the precise distance weighting used.  We believe the 
main source of systematic error that we have not accounted for is the 
incompleteness of the galaxy catalogue.  That is, some fraction of 
detectable GWs will originate in galaxies that are not included in the 
catalogue, and so the true host cannot be given a ranking.  
Our estimated probabilities for successful imaging should be multiplied 
by the catalogue completeness, estimated as 75\% to 50 Mpc and higher at 
smaller distances.  For comparison, our simulations reveal that 90\% of 
the GWs detectable by current instruments originate from galaxies within 
21 / 34 / 44 Mpc for 
1.4-1.4$M_{\odot}$ NS-NS / 1.4-5.0$M_{\odot}$ NS-BH / 1.4-10.0$M_{\odot}$ 
NS-BH binaries.

The approximation (equation (\ref{timing})) used for timing errors has been shown to 
underestimate the error for low-SNR signals by up to 20\% \cite{Fairhurst:2009tc}. 
Increasing the timing error for all detectors and all signals by 20\% was 
found to change the probabilities by only a few percent.  A more important 
factor is that our simulations treat the advanced Virgo detector as identical 
to advanced LIGO; using the design proposed in \cite{avirgo} 
reduces the probabilities by 5\%-10\% due to the lower distance sensitivity 
of the advanced Virgo design.

Finally, let us comment briefly on the applicability of our ranking statistic 
to the advanced LIGO \cite{aligo} and Virgo \cite{avirgo} detectors.  
Advanced LIGO will have maximum ranges of $\sim$450 Mpc for NS-NS systems 
and $\sim$930 Mpc for NS-BH systems.  This improved sensitivity presents two 
challenges for host identification: there are many more galaxies in a 
typical sky position error box; and we lack comprehensive galaxy catalogues 
to these distances.  While our technique appears to be promising for the 
most close-by binaries (within 100 Mpc, expected at a rate of a few per year), more 
extensive catalogues will be required to apply it to the majority of detected signals.
More generally, further investigation is needed of 
strategies for host galaxy identification in the advanced detector era.

\acknowledgments

The authors thank Jonah Kanner, Darren White, Richard O'Shaughnessy 
and Stephen Fairhurst for useful discussions.  We also thank Richard 
O'Shaughnessy for providing data on merger rates used in our simulations. 
Parts of this work were performed using the computational facilities of 
the Advanced Research Computing @ Cardiff (ARCCA) Division, Cardiff 
University, with assistance from Gareth Jones and James Osborne.
LKN was supported by a STFC studentship and PJS was supported in 
part by STFC grant PP/F001096/1. 
This document has been assigned LIGO Laboratory document
number {LIGO-P10}{00087}-v3.

\bibliography{references}

\end{document}